# SYNCHRONOUS HIGH-FREQUENCY DISTRIBUTED READOUT FOR EDGE PROCESSING AT THE FERMILAB MAIN INJECTOR AND RECYCLER

J.R. Berlioz*, M.R. Austin, J.M. Arnold, K.J. Hazelwood*, P. Hanlet, M.A. Ibrahim*,
A. Narayanan [1], D. J. Nicklaus, G. Praudhan, A.L. Saewert,
B.A. Schupbach, K. Seiya, R.M. Thurman-Keup, N.V. Tran,
Fermi National Accelerator Laboratory[†], Batavia, IL USA
J. Jang, H. Liu, S. Memik, R. Shi, M. Thieme, D. Ulusel,
Northwestern University[‡], Evanston, IL USA
[1]also at Northern Illinois University, DeKalb, IL USA

## Abstract

The Main Injector (MI) was commissioned using data acquisition systems developed for the Fermilab Main Ring in the 1980s. New VME-based instrumentation was commissioned in 2006 for beam loss monitors (BLM)[2], which provided a more systematic study of the machine and improved displays of routine operation. However, current projects are demanding more data and at a faster rate from this aging hardware. One such project, Real-time Edge AI for Distributed Systems (READS), requires the high-frequency, low-latency collection of synchronized BLM readings from around the approximately two-mile accelerator complex. Significant work has been done to develop new hardware to monitor the VME backplane and broadcast BLM measurements over Ethernet, while not disrupting the existing operations-critical functions of the BLM system. This paper will detail the design, implementation, and testing of this parallel data pathway.

## INTRODUCTION

The Real-time Edge AI for Distributed Systems (READS) project is a collaboration between the Fermilab Accelerator Division and Northwestern University. The project has two objectives: 1) to create a real-time beam loss de-blending system for the Main Injector (MI) and Recycler (RR) utilizing machine learning (ML) [3], and 2) to implement ML into the future Delivery Ring slow spill regulation system for the Mu2e experiment [4, 5]. This paper focuses on the creation of the data acquisition architecture, capable of data streams into both training sets and firmware implementation of ML models across a distributed network. The details of the ML model and the progress of disentangling the loss sources are not the topic of this paper but are described in detail within our conference sister paper [6].

---
* Equal paper contribution
† Operated by Fermi Research Alliance, LLC under Contract No.De-AC02-07CH11359 with the United States Department of Energy. Additional funding provided by Grant Award No. LAB 20-2261 [1]
‡ Performed at Northwestern with support from the Departments of Computer Science and Electrical and Computer Engineering

### BLM Data Collection

Beam loss monitors (BLMs) consist of a distributed network of VME front ends (or "nodes"), which captures spatially-identifiable and time-correlated ionizing radiation measurements from 259 argon-gas ionizing chambers, or BLM detectors, installed around the MI and RR. BLM systems report, on a logarithmic scale, losses for all BLM detectors at various times throughout a cycle.

To achieve this, charges measured from the BLMs detectors are integrated in 21 µs periods. These periods are used to construct three different sliding sums with user-defined time scales: 3 ms, 50 ms, and 1 s [7]. The sliding sums are transmitted via a VME crate controller card to ACNET, Fermilab's accelerator control system [8], for display and analysis. Simultaneously, these sums also drive primary inputs to a beam abort system, which compares the reported charge readouts and sliding sum values with beam loss abort threshold values for the MI [2].

Currently, the abort system struggles to distinguish and disentangle the source of the beam losses. This is because 1) MI and RR share an enclosure, 2) both machines can and do often have high-intensity beam in them simultaneously, and 3) both machines can generate significant beam loss. Accelerator operators use their expertise to determine the origin of a loss from a beam loss pattern. The process is not automatic, and the manual analysis creates unnecessary downtime. The READS project aims to deploy a ML model that will infer in real-time the machine loss origin.

Having training sets with the appropriate temporal and spatial resolution is paramount for the development of an accurate, realistic ML model. Unfortunately, the current BLM readout system presents a bottleneck for data collection and limits the reaction time for the ML model. Although BLM measurements update at 333 Hz (i.e., 3 ms integration period), the BLM readout system can only provide a maximum data rate of 30 Hz. Furthermore, the interface between the BLM and the abort system imposes operations-critical restrictions on any modifications to the existing BLM system. These reasons were the key motivations for developing an alternative solution through the VME Reader cards.

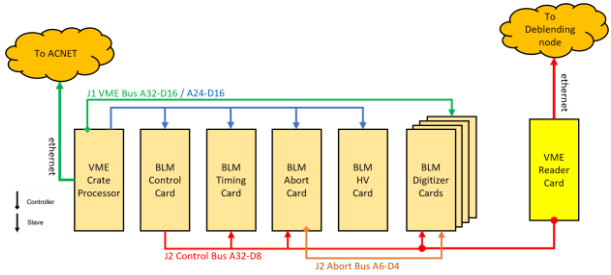

Figure 1: VME Reader card packet capture scheme.

## VME READER CARD

The VME Reader cards were designed to form a distributed network of data collection nodes. To achieve this goal, the VME Reader cards had to be incorporated into each BLM system. The BLM system is a VME-based architecture, composed of several cards. Each card was intentionally designed to handle a predefined subset of responsibilities within the system. More specifically, the Control card obtains and stores the sliding sums from the Digitizer cards in circular buffers, allowing the VME crate controller to read them at any time. Since this communication occurs across the VME J2 backplane, the VME Reader cards can monitor J2 activity for relevant information. Since the cards in the crate are agnostic to non-communicating cards connected to the backplane, the VME Reader card can collect the data for the ML model without interrupting the normal BLM operations. The VME Reader cards also maintain a dedicated Ethernet data link for the ML data collection. Figure 1 shows how the VME Reader card fits into the BLM system architecture.

### VME Reader Architecture

The workhorse of VME Reader cards is a MitySOM 5CSX System on Module (SOM) by Critical Link. A carrier board for the SOM was designed and assembled, which 1) mechanically fit within the existing VME architecture, 2) received appropriate clock events and machine state data from the control system, 3) provide an Ethernet communication link, and 4) provide the SOM access to the J2 connector.

In addition, the SOM comes with an Altera Cyclone V System on Chip (SOC), memory subsystems, and an on-board power supply. The FPGA fabric in the SOC was used to capture relevant, synchronized data sets from the VME-bus and then to transfer them to the dual-core ARM Cortex-A9, colloquially named by Altera/Intel as the Hard Processor System or HPS. Once in the HPS, the data is packaged and sent out through the Ethernet.

The system architecture was chosen to leverage the development framework and expertise used in other distributed system projects [9]. By utilizing a SOC (instead of an FPGA-only solution), version control as well as both firmware and software updates are made possible using U-boot and Yocto/Linux. Also, any software or firmware changes are pushed to the VME Reader cards remotely through the HPS. This setup allows for future scalability as well as rapid debug and support of the cards after deployment.

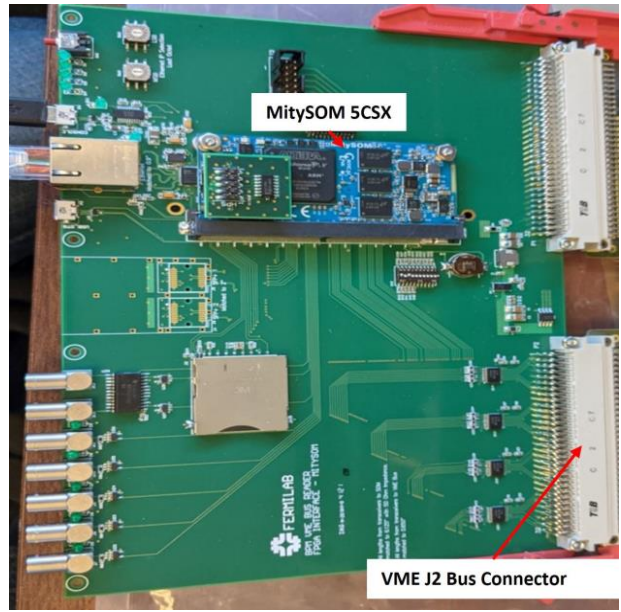

Figure 2: BLM VME Reader card layout.

### DDCP Streamer

The communication between the central ML node and distributed data collection node is established through a lightweight message protocol format, known as the Distributed Data Communication Protocol (DDCP) [10]. The DDCP framework was used to address control features in the VME Reader card, configuration of the data stream parameters, and enable the data stream.

A small collection and transmission time footprint is required for the VME Reader cards in the READS collection scheme. The combined time for data transmission after the data capture plus the time-of-travel on the wire, reception, and ML model activation was specified to be no higher than 3 ms by the READS project. The time-of-travel on the wire from the MI houses to the central node was measured to be on average 1 ms. Since the footprint limits the practicality of packet retransmission, the READS cards implemented a UDP-based DDCP streamer instead of a TCP-based one.

### Packet Structure

Each streamed packet from the VME reader cards consists of a DDCP header and a custom payload. Moreover, the payload contains of clock event information, the Fermilab accelerator complex's MI and RR machine state information, and the BLM sums read from the VMEbus. To synchronize the data streams, both epoch second and millisecond timestamps are used.

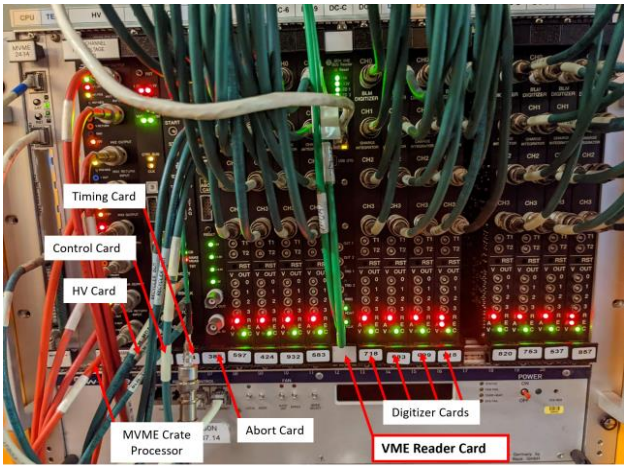

Figure 3: VME Reader card installed on BLM Crate

## DATA COLLECTION

The VME bus reader cards were all installed, tested and stream by early summer 2022. Figure 3 shows an installed VME Reader on a BLM VME Crate. At that time, data started to be collected for ML model training through a python script.

*Multi-Threaded Collection*

A python DDCP library was written for this project to help serialize and de-serialize packets. The resulting data collection is shown in Fig 4. To retrieve data from the VME reader cards, multi-threaded python code was written to collect the streamed UDP packets from all seven remote cards simultaneously. Each card's stream is monitored on a separate socket connection and port, and data that arrives on that port is pushed to its own synchronized queue. A monitoring thread frequently checks the card queues' size and structure to ensure the expected amount of data exists in each queue and that each data element is of the proper size. If the monitor thread finds any significant discrepancies with the data, warning messages are posted to a Slack channel used for monitoring data collection. Another thread is tasked with emptying each card's queue and writing the data to file at user-specified times of day, most often every five minutes on the minute.

The data stream has shown sporadic millisecond delay jitter not shown in Fig 4 for some packets during transmission. The jitter does not affect the data stream content as the code buffers delayed data before sending it out. The jitter is attributed to be a side effect of working with a non-RT version of Linux. Work is being done to replace the Linux system with one with the RT-patch included and optimize the streamer code.

## CONCLUSION AND FUTURE WORK

This paper demonstrates the implementation of a packet capturing node to create a dedicated channel on a VME-based BLM readout system. The system can capture data and successfully transmit packets over Ethernet with minimal cost and interference on the current BLM readout infrastructure.

Further READS work will focus on eliminating the transmission jitter from the cards and building the central node capable of making the inferences and handling the multiple data streams.

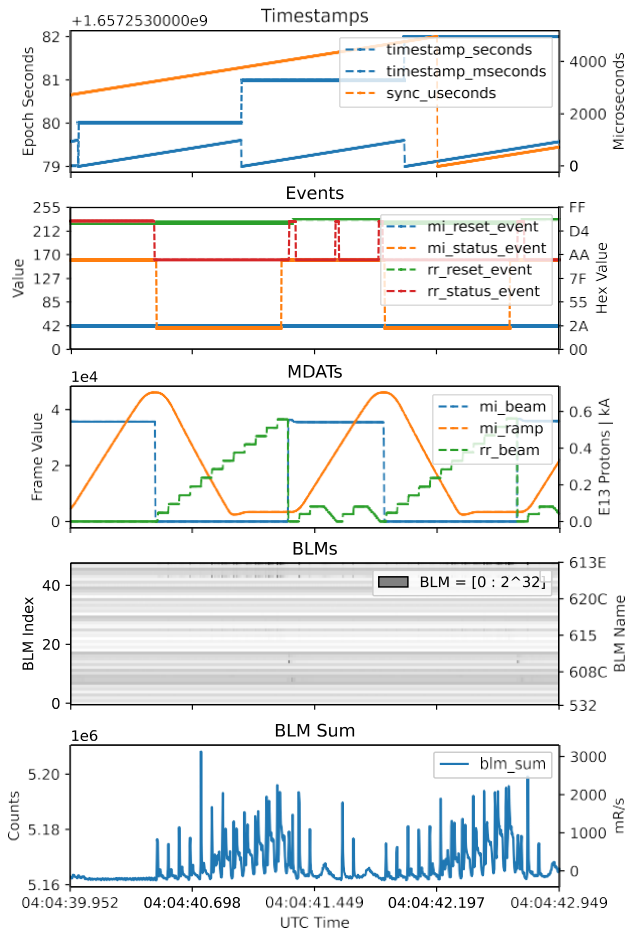

Figure 4: Three seconds of data from MI60S VME reader card

# REFERENCES

[1] Department of Energy, Office of Science. "Data, Artificial Intelligence, and Machine Learning at DOE Scientific User Facilities, DOE National Laboratory Program Announcement Number: LAB 20-2261." (2020), https://science.osti.gov/-/media/grants/pdf/lab-announcements/2020/LAB_20-2261.pdf

[2] A. Baumbaugh *et al.*, "The upgraded data acquisition system for beam loss monitoring at the fermilab tevatron and main injector," *Journal of Instrumentation*, vol. 6, no. 11, 2011, doi:10.1088/1748-0221/6/11/T11006

[3] K. Hazelwood *et al.*, "Real-Time Edge AI for Distributed Systems (READS): Progress on Beam Loss De-Blending for the Fermilab Main Injector and Recycler," in *Proc. IPAC'21*, Campinas, SP, Brazil, 2021, paper MOPAB288, pp. 912–915, doi:10.18429/JACoW-IPAC2021-MOPAB288

[4] L. Bartoszek *et al.*, "Mu2e Technical Design Report," 2014, doi:10.2172/1172555

[5] A. Narayanan *et al.*, "Optimizing Mu2e Spill Regulation System Algorithms," in *Proc. IPAC'21*, Campinas, Brazil, May 2021, pp. 4281–4284, doi:10.18429/JACoW-IPAC2021-THPAB243

[6] M. Thieme *et al.*, "Semantic Regression for Disentangling Beam Losses in the Fermilab Main Injector and Recycler," presented at NAPAC'22, Albuquerque, New Mexico, USA, Aug. 2022, paper MOPA28, unpublished.

[7] A. Baumbaugh *et al.*, *Beam Loss Monitor Upgrade Users' Guide*, 2010, https://beamdocs.fnal.gov/cgi-bin/sso/ShowDocument?docid=1410

[8] J. F. Patrick, "The Fermilab Accelerator Control System," in *Proc. ICAP'06*, Chamonix, Switzerland, Oct. 2006, pp. 246–249, https://jacow.org/icap06/papers/WEA2IS03.pdf

[9] J. S. Diamond and K. S. Martin, "Managing a Real-time Embedded Linux Platform with Buildroot," in *Proc. ICALEPCS*, 2015, p. 4, DOI:10.18429/JACoW-ICALEPCS2015-WEPGF096

[10] D. C. Voy, *DDCP Rule Book*, 2020, https://cdcvs.fnal.gov/redmine/projects/adinstsharedddcp/repository/revisions/master/entry/DDCPRulebook.xlsx